# Crystalline Phase Effects on the Nonlinear Optical Response of $MoS_2$ and $WS_2$ Nanosheets

*Michalis Stavrou,[a,b] Nikolaos Chazapis,[a,b] Eleni Nikoli,[c] Raul Arenal,[d,e,f] Nikos Tagmatarchis,[c] Stelios Couris[a,b]\**

[a] Department of Physics, University of Patras, 26504 Patras, Greece

[b] Institute of Chemical Engineering Sciences (ICE-HT), Foundation for Research and Technology-Hellas (FORTH), P.O. Box 1414, Patras 26504, Greece

[c] Theoretical and Physical Chemistry Institute, National Hellenic Research Foundation, 48 Vassileos Constantinou Avenue, 11635, Greece

[d] Laboratorio de Microscopias Avanzadas (LMA), Universidad de Zaragoza, Mariano Esquillor s/n, 50018 Zaragoza, Spain

[e] Instituto de Nanociencia y Materiales de Aragon (INMA), CSIC-U. de Zaragoza, Calle Pedro Cerbuna 12, 50009 Zaragoza, Spain

[f] ARAID Foundation, 50018 Zaragoza, Spain

**Corresponding Author**

\* Stelios Couris (email: couris@iceht.forth.gr, couris@upatras.gr)

**ABSTRACT**

In the present work, some $MoS_2$ and $WS_2$ nanosheets were prepared and characterized. Depending on the preparation procedures, trigonal prismatic (2H) or octahedral (1T) coordination of the metal atoms were obtained, exhibiting metallic (1T) or semiconducting (2H) character. Both $MoS_2$ and $WS_2$ nanosheets were found exhibiting large nonlinear optical (NLO) response, strongly dependent on their metallic (1T) or semiconducting (2H) character. So, the semiconducting character $2H-MoS_2$ and $2H-WS_2$ exhibit positive nonlinear absorption and strong self-focusing, while their metallic character counterparts exhibit strong negative nonlinear absorption and important self-defocusing. In addition, the semiconducting $MoS_2$ and $WS_2$ were found exhibiting important and very broadband optical limiting action extended from 450 to 1750 nm. So, by selecting the crystalline phase of the nanosheets, i.e., their semiconduction/metallic character, their NLO response can be greatly modulated. The results of the present work demonstrate unambiguously that the control of the crystalline phase of $MoS_2$ and $WS_2$ provides an efficient strategy for 2D nanostructures with custom made NLO properties for specific optoelectronic and photonic applications.

**INTRODUCTION**

The discovery of graphene in recent years has been a significant milestone for the scientific community, paving the way to numerous applications stemming from its unique physicochemical and optoelectronic properties.[1-3] Simultaneously, the resounding success of graphene has opened up a new avenue for the development of alternative two-dimensional layered materials (2DLMs). A rapidly emerging class of such 2DMLs which has demonstrated great prospects in a wide range of practical applications, are the transition metal dichalcogenides (TMDCs), whose generalized chemical formula is $MX_2$, where M refers to a transition metal atom (e.g., Mo, W, Nd) and X represents a chalcogen atom (e.g., S, Se, Te). Currently, there have been intense research efforts towards the applications of these 2DLMs in various optoelectronic devices, as e.g., in field effect transistors,[4] solar cells,[5] efficient photodetectors,[6] photoelectric modulators,[7] passive mode-lockers,[8] optical switchers,[9] for drug delivery,[10,11] biosensing,[12,13] and several others.

In contrast to graphene's individual atom-thick layer(s), TMDCs are composed of a sandwich-like structure of a transition metal atoms-layer between two layers of chalcogen atoms. Depending on the atomic stacking configuration, layered TMDCs can form two common thermodynamically stable structural phases characterized by either trigonal prismatic (2H) or octahedral (1T) coordination of the metal atoms.[14] It is noteworthy that each phase can be easily turned into the other through intralayer atomic plane gliding. In that direction, 2H-$MoS_2$, for instance, can be transformed to 1T-$MoS_2$ by Li or K intercalation.[15,16] The diversity of crystalline structure and structural phase of the d electrons, as well as the number and type of layer stacking sequences, result in a broad range of electronic properties, from the point of view of the band structure character of these layered materials. For example, octahedral phase (1T) $MoS_2$, $MoSe_2$, $WS_2$, and $WSe_2$ are metals, while in their 2H structural phase they are semiconductors, compensating, at large degree, the weakness of zero-bandgap graphene in optoelectronics. Additionally, TMDCs exhibit a thickness-dependent transition from indirect bandgap semiconductor bulk material to direct bandgap semiconductor single-layer. For example, $MoS_2$ and $WS_2$ possess a direct bandgap of ~1.8 eV in their single-layered configuration while they exhibit an indirect bandgap of ~1.3 eV in their bulk form.[17,18] In particular, the latter immediately imparts some unique properties in TMDCs, such as high carrier mobility and exceptional nonlinear absorptive response.[18,19]

The strong optical nonlinearities of 2DMLs provide a valuable tool for realizing different nonlinear optical (NLO) processes in photonic and optoelectronic devices, such as optical switching,[20] optical frequency conversion,[21] optical limiting,[22-24] and saturable absorption.[25] Other classes of 2DMLs, beyond graphene, as e.g., the $MoS_2$ and $WS_2$, fulfill the requirements of ideal nonlinear optical materials, exhibiting strong and ultrafast NLO response, broadband optical absorption, large damage threshold and high chemical stability. So far, many research articles have been published concerning the NLO properties of semiconducting $MoS_2$ and $WS_2$, revealing exceptional broadband saturable absorption (SA),[26,27] significant optical limiting (OL) efficiency,[28] and size-dependent NLO response.[29] However, these studies are limited to the semiconducting 2H crystalline phase of $MoS_2$ and $WS_2$, while a comprehensive comparison of the NLO response of the semiconducting and metallic phases of $MoS_2$ and $WS_2$ is rather scarce. Furthermore, the studies on the OL performance of semiconducting $MoS_2$ and $WS_2$ are limited to wavelengths up to 1064 nm, thus narrowing the range of their applications in OL devices. In that view, the current work aims to elucidate the effect of crystalline phase on the NLO properties and the broadband OL performance of $MoS_2$ and $WS_2$. In that view, both semiconducting (2H-$MoS_2$ and 2H-$WS_2$) and metallic (1T-$MoS_2$ and 1T-$WS_2$) phase $MoS_2$ and $WS_2$ samples were prepared and characterized. Then, their NLO properties were investigated using 4 ns, visible (532 nm) and infrared (1064 nm) laser excitation, while their optical limiting performance was studied in detail in a broadband spectral region, extending from 450 to 1750 nm.

**EXPERIMENTAL SECTION**

**Exfoliation of TMDCs**

For obtaining the 2H-semiconducting phase of $MoS_2$ and $WS_2$, 150 mg of bulk TMDCs ($MoS_2$ or $WS_2$) were mixed with 15 mL of chlorosulfonic acid and the suspension was sonicated for 8 h and stirred vigorously at room temperature for a week. Then, the suspension was dropwise added to 200 mL of distilled water, under an ice bath, and it was filtered on PTFE filter (pore size 0.1 μm) and extensively washed with distilled water. The solid residue was added to 200 mL of N-methyl pyrrolidone (NMP) and ultra-sonicated for 1.5 h (tip sonication at 40% of amplitude 100% of 200W). The suspension was left overnight to settle, and the supernatant was then collected by pipette and filtrated on PTFE filter (0.2 μm pore size), washed with a large amount of methanol, acetone and dichloromethane, and dried in vacuum.

For obtaining the 1T-metallic phase, bulk $2H-MoS_2$ or $2H-WS_2$ (750 mg, 4.68 mmol) was heated at 200 °C for 24h and after cooling the TMDCs powder to room temperature, the bulk material was placed under nitrogen atmosphere in a flame-dried round-bottom flask equipped with a magnetic stir bar. Then, 2.5M n-BuLi (7.5 mL, 18.75 mmol) was introduced, and the mixture left under vigorous stirring for 48h under nitrogen. After that period, the mixture was left to settle, separated in two phases and carefully the supernatant was removed by pipette under nitrogen. Then, dry hexane (10 mL) was added, the mixture stirred for 10 mins and this process was repeated twice in order to remove the remained n-BuLi. After that, the suspension was rapidly immersed in distilled water (750 mL) under ice bath and the mixture was ultra-sonicated for 3h (tip sonication at 50% of amplitude 100% of 200W, pulse ON: 5s – OFF: 5s) under ice bath and left to settle overnight. Then, the 2/3 of the supernatant was pipetted and stored in dark at room temperature in sealed flask. The concentration of the suspension was calculated at 1.2 mg/mL. Particularly, for obtaining the metallic $1T-WS_2$, a heating treatment under nitrogen at 68 ºC was employed.

**TEM imaging**

High-resolution scanning transmission electron microscopy (STEM) imaging has been developed in a probe-corrected Thermo Fisher Titan-Low-Base 60-300 operating at 120 kV (equipped with a Cs-probe corrector (CESCOR from CEOS GmbH)). For performing these TEM studies, the samples have been dispersed in ethanol and the suspensions have been ultrasonicated and dropped onto copper carbon holey grids.

**NLO and OL measurements**

1 mg powder of 2H-MoS$_2$, 1T-MoS$_2$, 2H-WS$_2$, and 1T-WS$_2$ was dispersed in 1 mL of different solvents (e.g., water, dimethylformamide, acetonitrile, toluene, etc.). The dispersions were prepared in vials after 30 min sonication of the TMDCs powder, and they were left to rest for a few days to check the stability of the as-prepared dispersions concerning precipitation. However, only the 1T-MoS$_2$ and 1T-WS$_2$ aqueous dispersions and the dimethylformamide (DMF) 2H-MoS$_2$, and 2H-WS$_2$ dispersions were chosen for the NLO and OL measurements, as they exhibited a greater stability.

At first, the impact of the crystalline phase of the TMDCs dispersions on their NLO properties was investigated by means of the Z-scan technique.[30] Z-scan technique is a relatively simple technique which is widely used for the characterization of the third-order optical nonlinearities of materials (dispersions, films, etc.). Z-scan was chosen as it allows the simultaneous determination of the sign and the magnitude of the nonlinear absorption coefficient β, and the nonlinear refractive index parameter γ′, from a single measurement. Briefly, the Z-scan experimental procedures consist basically of measuring the variation of the normalized transmittance of a sample, as it moves along the propagation direction (z-axis) of a focused laser beam, thus experiencing variable incident laser intensities at each z-position. As the sample approaches the focal plane of the laser beam (i.e., at z=0), where the intensity becomes sufficiently high to induce the NLO absorption and refraction of the sample, the sample's transmittance characteristics are modified. The previously mentioned NLO quantities are expressed in terms of the nonlinear absorption coefficient β and the nonlinear refractive index parameter γ′, respectively. Then, the values of the nonlinear absorption coefficient β and the nonlinear refractive index parameter γ′ can be evaluated using two different experimental configurations, the so-called "open-aperture" (OA) and "closed-aperture" (CA) Z-scans. In the former Z-scan transmission measurement, the laser light which is transmitted through the sample is totally collected by a lens and is measured e.g., by a photomultiplier or a photodiode, providing information about the NLO absorption of the sample. In the latter transmission measurement (i.e., the CA Z-scan) the transmitted laser beam passes through a narrow pinhole positioned in the far-field and then it is measured by a photomultiplier or a photodiode and provides information on the sample's nonlinear refraction. The signs of β and γ′ can be deduced from the shapes of the OA and CA Z-scan recordings. So, the presence of a transmission maximum/minimum in the OA Z-scan denotes saturable absorption/reverse saturable absorption, i.e., SA corresponding to β<0, or RSA corresponding to β>0, respectively. Correspondingly, the presence of a pre-focal transmission minimum followed by a post-focal maximum or *vice versa* in the CA Z-scan recording denotes self-focusing (γ′>0) or self-defocusing (γ′<0) behavior, respectively. When there is significant NLO absorption, to remove its influence on the corresponding (simultaneously measured) CA Z-scan, the latter is divided by the corresponding OA Z-scan. The resulting Z-scan called "Divide" Z-scan, allows the accurate determination of the magnitude of the nonlinear refractive index parameter γ′.

The nonlinear absorption coefficient β and the nonlinear refractive index parameter γ′ are deduced by fitting the experimental OA and "divided" Z-scans with Eq. (1) and Eq. (2), respectively,[30,31]

$$T(x) = \frac{1}{\sqrt{\pi}(\frac{\beta I_0 L_{eff}}{1+[x]^2})} \int_{-\infty}^{+\infty} \ln[1 + \frac{\beta I_0 L_{eff}}{1+[x]^2} e^{-t}]\, dt \quad (1)$$

$$T(x) = 1 - \frac{4\gamma' k I_0 L_{eff} x^2}{(1+x^2)(9+x^2)} \quad (2)$$

where $x=z/z_0$ with z and $z_0$ denoting the sample's position and Rayleigh length, respectively, $I_0$ is the on-axis peak irradiance, $L_{eff}$ is the sample's effective length and k is the wavenumber at the excitation wavelength.

Z-scan experiments were performed at two excitation wavelengths, using the fundamental (at 1064 nm) and the second harmonic (at 532 nm) outputs from a 4 ns Q-switched Nd:YAG laser operating at a repetition rate from 1 to 10 Hz. The laser beam was focused by means of a 20 cm focal length quartz planoconvex lens into a 1 mm thick (*l*) quartz cell, containing the TMDCs dispersions. The laser beam waist at the focal plane was measured using a CCD camera. It was found to be (18 ± 4) and (30 ± 4) μm at 532 and 1064 nm, respectively. The corresponding Rayleigh length $z_0$ was calculated to be 1.91 and 2.65 mm, at 532 and 1064 nm, respectively, thus satisfying the thin sample approximation requirement, i.e., *l*<$z_0$, that is necessary for the validity of the Z-scan approximations For the accurate determination of the NLO response of the $MoS_2$ and $WS_2$ samples, systematic Z-scan experiments using different concentration dispersions were performed, ranging from 0.05 to 0.3 mg/mL, at various incident laser intensities.

To study the optical limiting, OL, performance of the TMDCs dispersions, transmission measurements were conducted over a wide spectral range, from 450 to 1750 nm. For these measurements, the 1064 and 532 nm outputs of the Nd:YAG laser described previously and the output from an optical parametric oscillator (OPO) unit pumped by the same Nd:YAG laser were used. More specifically, measurements were performed at 450, 532, 650, 750, 850, 950, 1064, 1150, 1250, 1350, 1450, 1550, 1650 and 1750 nm. For the OL

measurements, the cells containing the dispersions of the TMDCs samples were placed at the focal plane of a 20 cm focal length plano-convex lens and were irradiated with different energy laser pulses, while the incident and transmitted laser energies were monitored by two identical calibrated joulemeters, just before and after the cell. Then, the input and output energy fluencies were calculated considering the beam radii at focus, for the different irradiation wavelengths. The beam radii at focus were measured using a CCD camera and were found to be ~18-23 μm for visible laser wavelengths, and ~30-42 μm for NIR laser wavelengths.

During the measurements of the NLO response and the OL, several precautions were taken. So, the prepared dispersions were routinely stirred and spectrophotometrically checked before and after irradiation to ensure their stability, by taking their UV-Vis-NIR absorption spectra using a double beam spectrophotometer. Furthermore, to minimize or prevent the presence of undesirable cumulative thermal effects, the repetition rate of the laser was set at 1 Hz. In addition, the presence of nonlinear scattering (NLS), which could contribute to the OL action, was checked as well. For that purpose, a sensitive photodiode placed on a goniometric table behind the sample cell was used, monitoring the scattered laser radiation at different angles with respect to the laser beam propagation axis. For the range of incident laser intensities used, the NLS signal was very weak, indicating a rather negligible contribution on the OL performance of the studied samples.

**RESULTS AND DISCUSSION**

Different liquid-phase exfoliation protocols were followed to yield selectively the semiconducting 2H-MoS$_2$ and WS$_2$ and metallic 1T-MoS$_2$ and WS$_2$ materials, as depicted in Figure 1. Briefly, for obtaining the semiconducting phase of MoS$_2$ and WS$_2$, chlorosulfonic acid molecules were inserted between the layers of the nanosheets, keeping them apart due to developed electrostatic repulsive forces. Then, addition of water results in thermal decomposition of the superacid, breaking the van der Waals forces between the layers and contributing to the exfoliation.[32] On the other hand, upon n-BuLi treatment of MoS$_2$ and WS$_2$ results in transfer of electrons on the basal plane of the nanosheets, which are stabilized by lithium cations via intercalation.[15,33] The violent decomposition of the so-formed Li$_x$-MoS$_2$ and Li$_x$-WS$_2$ in aqueous media affords stable suspensions of nanosheets, in which the Mo or W atoms respectively, are octahedrally (1T) coordinated.

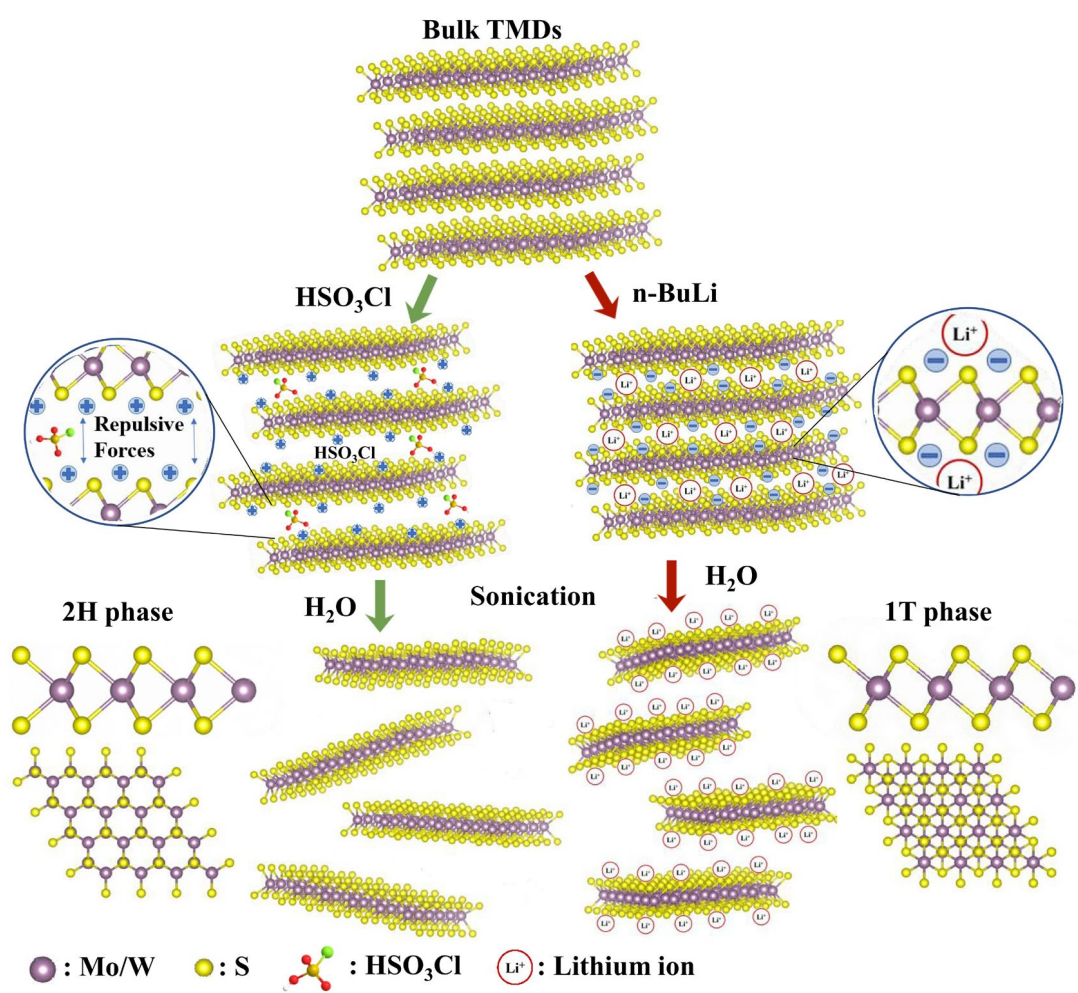

**Figure 1.** Exfoliation procedures for obtaining 2H-semiconducting and 1T-metallc MoS$_2$ and WS$_2$ samples.

Insights into the 2H and 1T-phase of exfoliated MoS$_2$ and WS$_2$ were obtained from Raman spectroscopy. Particularly, resonance Raman spectroscopy for exfoliated 2H-MoS$_2$, with respect to the energy of the A exciton (633 nm excitation), and for exfoliated 2H-WS$_2$, with respect to the energy of the B exciton (514 nm excitation), shows intense modes. By adjusting the laser power to 0.3 mW/cm$^2$ and with a short exposure time of 10 s to avoid sample overheating as well as to guarantee stability of the samples during spectral acquisition, Raman-active (A$_{1g}$-LA(M)) at 178 cm$^{-1}$, E$_{2g}^1$ at 375 cm$^{-1}$, A$_{1g}$ at 405 cm$^{-1}$ and 2LA(M) at 450 cm$^{-1}$ modes for 2H-MoS$_2$ are observed (Figure 2a). Similarly, the corresponding Raman bands, 2LA(M) at 348 cm$^{-1}$, E$_{2g}^1$ at 353 cm$^{-1}$ and A$_{1g}$ at 418 cm$^{-1}$, for 2H-WS$_2$, are shown in Figure 2b.[34] However, the Raman spectra for the metallic 1T-MoS$_2$ and 1T-WS$_2$ are richer in features, especially at the low frequency region. Briefly, on top of the aforementioned in-plane E$_{2g}^1$ and out-of-plane A$_{1g}$ modes, the characteristic J$_1$, J$_2$ and J$_3$ phonon modes,[35,36] located at 150, 220 and 328 cm$^{-1}$ for 1T-MoS$_2$ (Figure 2a) and at 130, 211 and 280 cm$^{-1}$ for 1T-WS$_2$, respectively, are evident (Figure 2b). The Raman spectra shown in Figure 2 represent the average of 120 measurements performed at different spots of different flakes of the materials, with the original spectra shown in the supporting information (Figure S1).

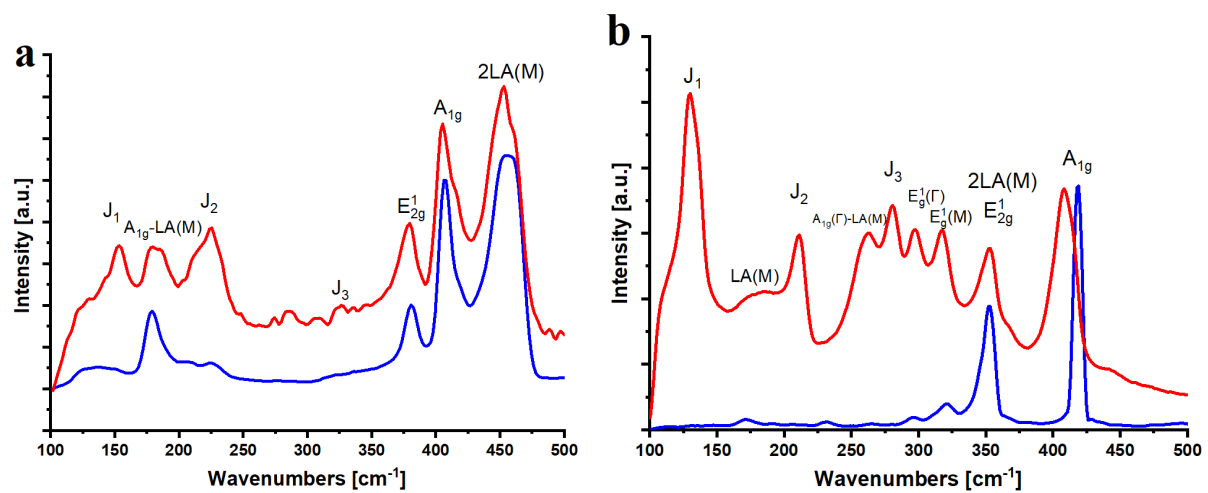

**Figure 2.** Raman spectra (average of 120 measurements) for **(a)** 2H-MoS$_2$ (blue) and 1T-MoS$_2$ (red) upon 633 nm excitation, and **(b)** 2H-WS$_2$ (blue) and 1T-WS$_2$ (red) upon 514 nm excitation, obtained at ambient conditions.

High-resolution scanning transmission electron microscopy (HRS-TEM) imaging is a very powerful technique for investigating the structure of this kind of nanomaterials.[32,37-39] The two different structural phases (2H and 1T) of these two kinds of transition metal dichalcogenide materials ($MoS_2$ and $WS_2$) are depicted in Figure 3. These high-angle annular dark-field (HAADF) HRS-TEM micrographs show and confirm the clear differences between the 2H and 1T phases of these samples.[32,37,38]

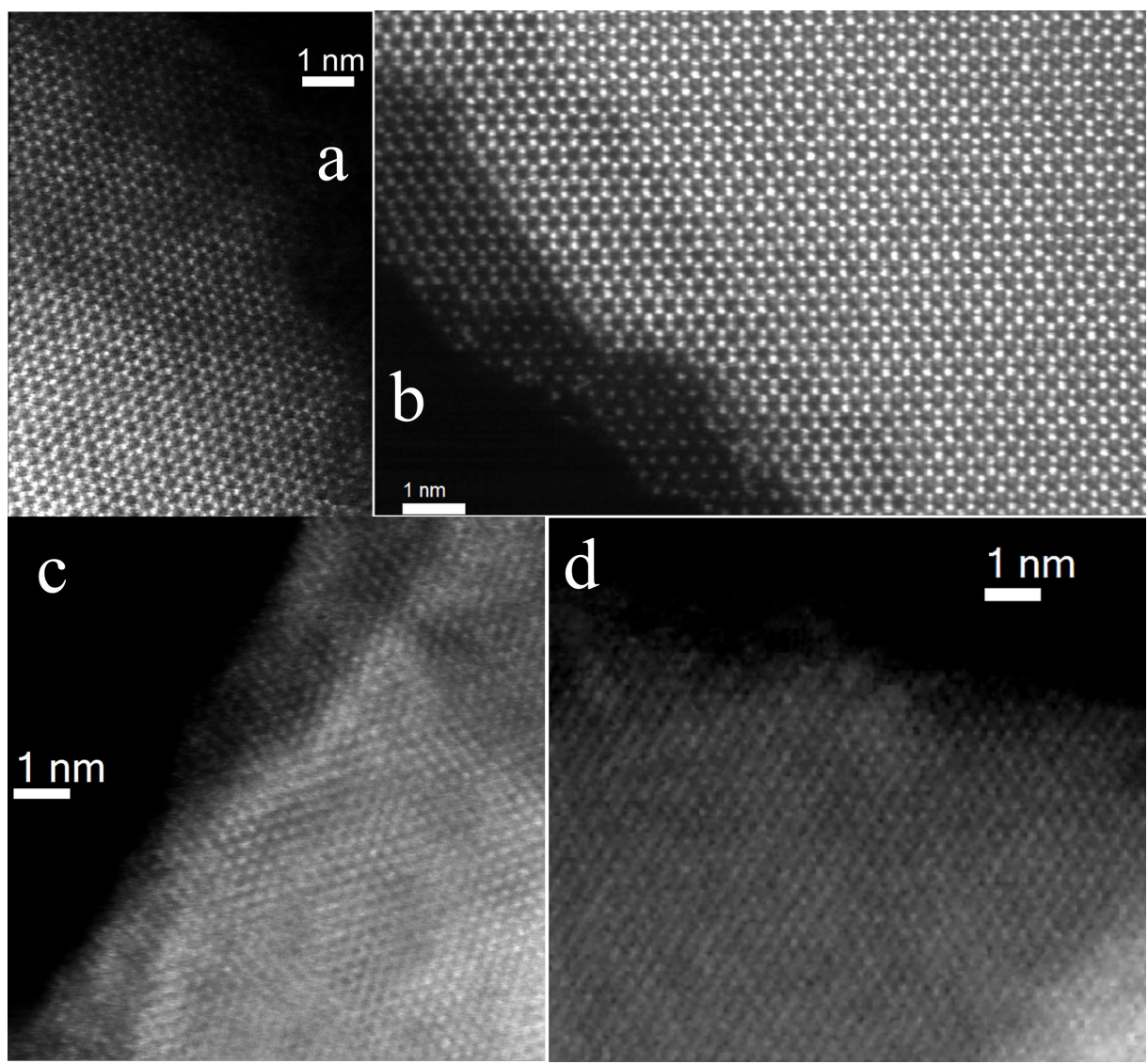

**Figure 3.** High-resolution scanning TEM HAADF images of 2H-$MoS_2$ and 2H-$WS_2$ (a and b) and 1T-$MoS_2$ and 1T-$WS_2$ (c and d).

In Figure 4, some representative UV-Vis-NIR absorption spectra of trigonal prismatic and octahedral phase $MoS_2$ and $WS_2$ nanosheets are presented, all having a concentration of 1 mg/mL. As shown, the absorption spectrum of 2H-$MoS_2$ exhibits two distinct peaks located at ~617 and 679 nm, ascribed to the direct excitonic transitions from the spin-orbit split valence band to the conduction band at the K point of Brillouin zone, namely the B and A excitons, respectively.[40,41] The broad band centered at ~427 nm emerges from excitonic transitions between the high-density states of valence band and conduction band at the point M of Brillouin zone.[40,41] On the other hand, 1T-$MoS_2$ displays some high energy excitonic features, located at the UV spectral region, i.e., at ~300 and 370 nm, indicating a phase conversion upon lithium exfoliation. Similarly, 2H-$WS_2$ exhibited two characteristic excitonic absorption peaks at ~532 and 645 nm, attributed to the B and A excitons, respectively. The excitonic transitions between the density of states peaks between the valence band and conduction band give rise to the band at ~460 nm.[42] Finally, 1T-$WS_2$ reveals excitonic transitions both at the UV and visible regions, as indicated by the shoulders appearing at 250 and 530 nm, respectively.

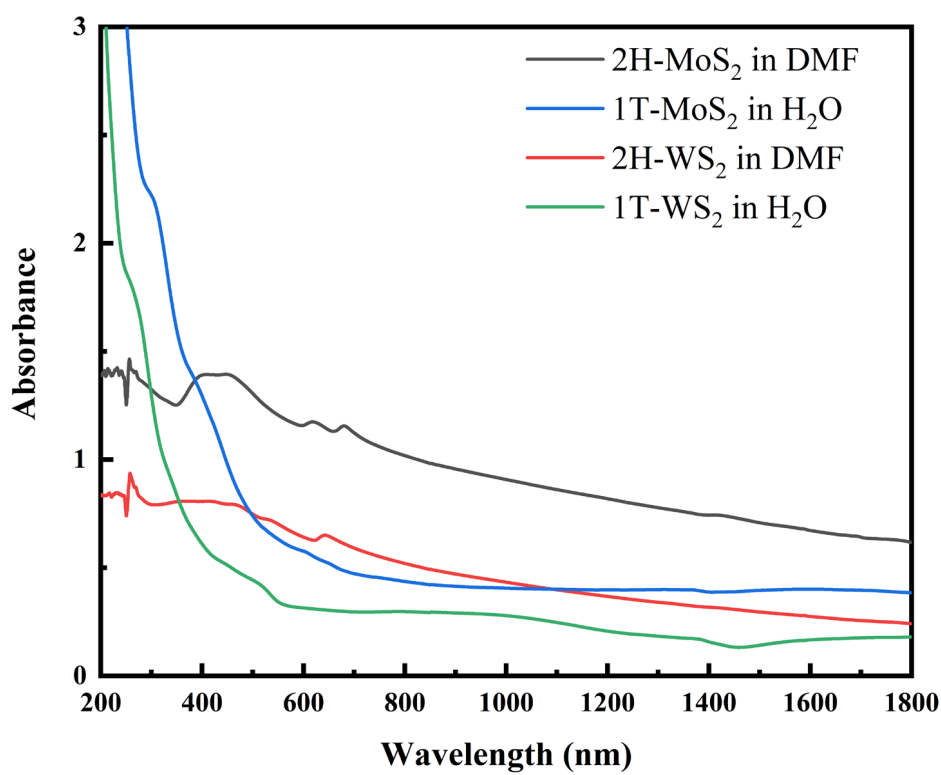

**Figure 4.** UV-Vis-NIR absorption spectra of 2H-$MoS_2$, 1T-$MoS_2$, 2H-$WS_2$, and 1T-$WS_2$ (all corresponding to a concentration of 1 mg/mL).

In Figures 5 and 6, some representative OA Z-scans of the 2H-MoS$_2$, 1T-MoS$_2$, 2H-WS$_2$, and 1T-WS$_2$ dispersions obtained under 532 and 1064 nm laser excitation, using different laser intensities, are presented. To facilitate comparisons, all dispersions had a concentration of 0.1 mg/mL. To check for possible contribution of the solvents used (i.e., water, DMF) to the NLO response of the dispersions, separate Z-scan measurements of the neat solvents were performed under identical experimental conditions to those used for the NLO measurements of the TMDCs dispersions. These measurements have shown that the solvents exhibited negligible NLO response; therefore, the OA Z-scans of the TMDCs dispersions of Figures 5 and 6 reveal directly the NLO absorptive response of the TMDCs. Moreover, as can be seen from Figures 5a,b and 6a,b, the OA Z-scans of the aqueous dispersions of the 1T-MoS$_2$ and 1T-WS$_2$ were found presenting a transmission maximum around the focal plane (i.e., at z=0), under both excitation wavelengths, indicative of a typical saturable absorption (SA) behavior. In addition, this transmission maximum was found to increase with incident laser intensity, remaining, however, a clear transmission maximum even for the highest laser intensity employed, suggesting that one-photon absorption (1PA) is the dominant process contributing to the NLO absorption in the case of metallic phase (1T-) MoS$_2$ and WS$_2$ samples. In contrast, the OA Z-scans of the DMF dispersions of 2H-MoS$_2$ and 2H-WS$_2$ (see Figure 5c,d and 6c,d) were found to exhibit a SA behavior only at low incident laser intensities, transformed gradually to a reverse saturable (RSA) behavior at higher laser intensities, as indicated by the increasing of the dip with the laser intensity. This behavior is typical of a transition from SA to RSA behavior. In fact, for high enough laser intensity, a clear RSA behavior was achieved.

Considering the interplay between SA and RSA, the magnitude of the nonlinear absorption coefficient β of the samples can be determined by adopting the modified form of the intensity-dependent absorption coefficient α(I):[43,44]

$$a(I) = \frac{a_0}{1+I/I_s} + \beta I \qquad (3)$$

where the first term is related to a negative sign absorptive nonlinearity (corresponding to SA), while the second term is related to a positive sign absorptive nonlinearity (corresponding to RSA or two-photon absorption (2PA)). Specifically, β is the nonlinear absorption coefficient corresponding to RSA or 2PA, occurring at relatively high laser intensity, while $I_s$, i.e., the saturation intensity, is defined as the intensity at which the absorbance of the sample is reduced to half of its linear value $a_0$ due to saturation (i.e., SA behavior). It is noteworthy that the nonlinear absorption coefficient β is a concentration-dependent parameter, while the saturation intensity $I_s$, in general, does not depend on the concentration. Then, the normalized transmittance, T(z), of an OA Z-scan can be written in the following form:

$$T(z) = \sum_{m=0}^{\infty} \frac{\left[-\frac{\alpha I_0 L_{eff}}{1+z^2/z_0^2}\right]^m}{m+1} \qquad (4)$$

By substituting eq. (3) in eq. (4), the parameters $I_s$ and β can be deduced by fitting the OA Z-scan recordings with the resulted expression. In Figures 5 and 6, some representative examples of such fittings are presented, where, as can be seen, an excellent agreement was found between the experimental data (full points) and the fitting curves (solid lines). Following this procedure, the values of the nonlinear absorption coefficient β for the different concentration dispersions were determined and are listed in Tables S1 and S2. As expected, the saturation intensity $I_s$ was found to be practically concentration-independent. In particular, the saturation intensity $I_s$, for 2H-$MoS_2$ and 2H-$WS_2$ samples, was determined to be about 0.043 and 0.075 GW/cm$^2$ at 532 nm, respectively, and about 0.040 and 0.060 GW/cm$^2$ at 1064 nm.

For the better understanding of the physical processes responsible for the NLO absorptive response of the trigonal prismatic and the octahedral phases of $MoS_2$ and $WS_2$, the electronic transitions occurring in these layered 2D nanostructures should be considered. So, the relatively smooth optical absorption profiles of the UV-Vis-NIR spectra (see e.g., the visible to NIR spectral region shown in Figure 4), suggest that the occurring interband optical transitions should be of resonant character, associated with 1PA process. As a result, upon 532 and/or 1064 nm irradiation, valence band electrons can be excited to the relatively high energy levels of the conduction band only under sufficient laser intensity. In this case, the photo-generated electron-hole pairs are converted to hot carriers. The latter are cooling down shortly, after redistributing their energy among carriers having lower energies, through two different mechanisms, i.e., carrier-carrier and electron-phonon scattering. The former relaxation process has been reported to occur at ~2 and ~1.4-2.4 ps after excitation for the 2H-$MoS_2$ and the 2H-$WS_2$, respectively, while the latter occurs later, i.e., after ~34 and ~15.8-46.6 ps for the 2H-$MoS_2$ and 2H-$WS_2$, respectively.[45,46] Thus, the photo-excited carriers relax and populate the bottom of valence and conduction bands, building an equilibrium electron-hole distribution. Increasing the incident laser intensity, the number of excited carriers accumulated in the conduction band is progressively increasing, resulting in the depletion of all the empty band states, thus impeding further excitation of carriers, owing to the Pauli exclusion principle. This behavior, known as Pauli blocking, results in the absorption bleaching of the samples, which is expressed by the manifestation of SA behavior. This is evidenced by the OA Z-scans shown in Figures 5 and 6. For the case of the 2H-MoS$_2$ and 2H-WS$_2$ samples, all were found exhibiting SA behavior at low laser intensities, due to the Pauli blocking mechanism. However, at higher laser intensities, the SA response turns to RSA response, due to two- or multi-photon absorption (2PA or MPA) processes which become possible only at higher laser intensity levels.[47] Again, the OA Z-scans shown in Figures 5 and 6, demonstrate nicely this passage from SA to RSA, and the gradual development of the transmission dip, in the centre of the OA Z-scan (where the maximum irradiance occurs during the z-scan experiment).

To further investigate the multiphoton absorption processes occurring at the higher intensity regime, i.e., when RSA occurs, the OA Z-scan recordings of 2H-MoS$_2$ and 2H-WS$_2$ were fitted using eq. (5), which is a more general form of eq. (4) describing the intensity-dependent absorption coefficient α(I):

$$a(I) = \frac{a_0}{1+(I/I_s)^{n-1}} + \beta_n I^{n-1} \qquad (5)$$

where n takes integer values greater or equal than 2 (i.e., n=2, 3…), corresponding to two-, three-, or n-photon process. In all cases, the equation 3, corresponding to a 2PA process, was found to fit better the experimental data of all studied TMDCs, for the range of laser intensities employed. Therefore, it can be safely concluded that the NLO absorptive response of 2H-MoS$_2$ and 2H-WS$_2$ at high laser intensities can be well described considering 2PA process. It is worth to add that the presence of any free-carrier absorption or excited state absorption (ESA) can be considered as negligible, since the deduced nonlinear absorption coefficient β under different laser intensities was found unchanged.[48]

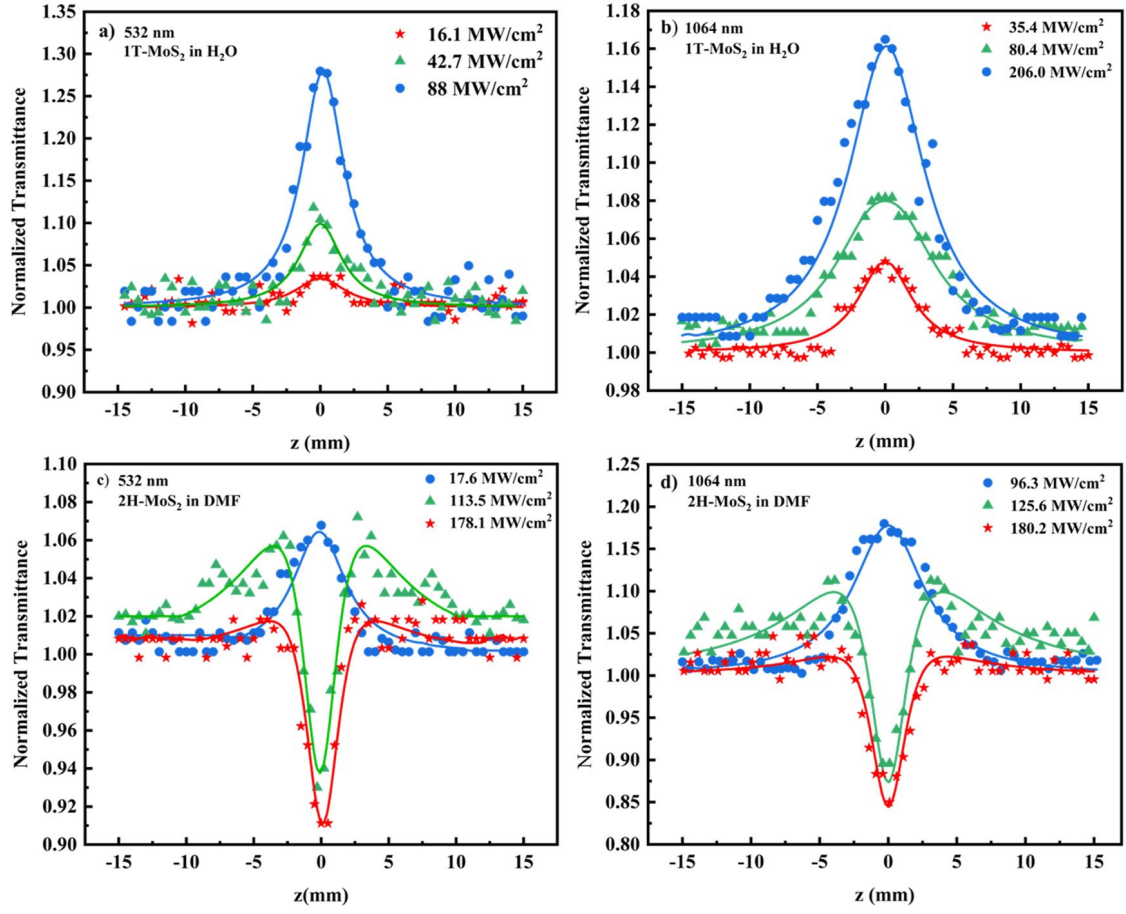

**Figure 5.** OA Z-scans of 1T-MoS$_2$, 2H-MoS$_2$ dispersions, under different laser excitation intensities at (a, c) 532 and (b, d) 1064 nm. All dispersions had a concentration of 0.1 mg/mL.

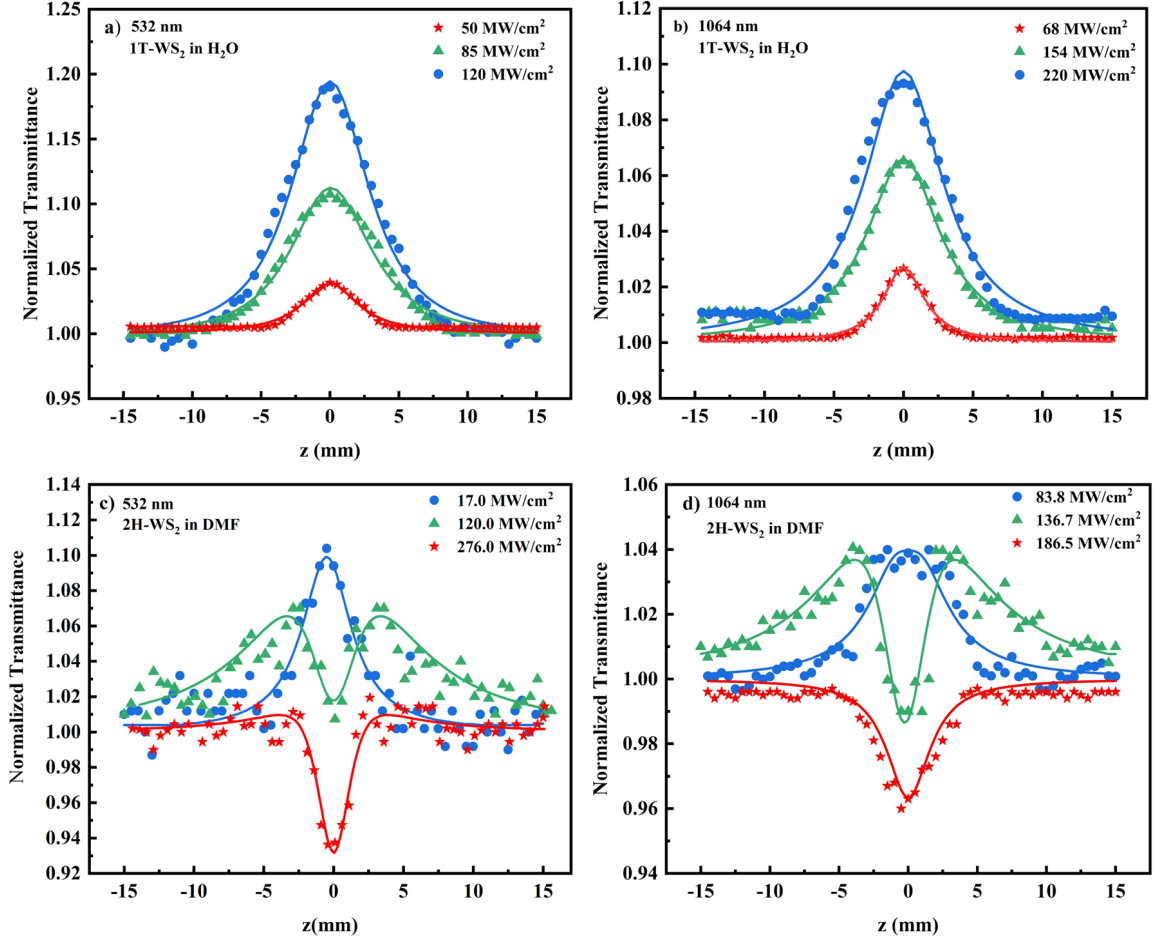

**Figure 6.** OA Z-scans of 1T-WS$_2$ and 2H-WS$_2$ dispersions, under different laser excitation intensities at (a, c) 532 and (b, d) 1064 nm. All dispersions had a concentration of 0.1 mg/mL.

In a recent work, studying the NLO absorption of some TMDCs nanosheets, including among others some semiconducting MoS$_2$, MoSe$_2$, WS$_2$, and WSe$_2$, their NLO absorption was measured using a 6 ns, 532 and 1064 nm laser excitation at low repetition rate (i.e., 2 Hz), i.e., very similar excitation conditions to those used in this work.[28] The MoS$_2$ and WS$_2$ were reported to exhibit saturable absorption at low laser intensities, converted to reverse saturable absorption at higher intensities, similarly to what was observed in the present study. The transition from SA to RSA was attributed to the combined action of a 2PA process, operating effectively at higher laser intensities, and to nonlinear scattering (NLS), attributed to laser-induced micro-bubbles and micro-plasma, both behaving as scattering centers.[28] In addition, RSA due to NLS has been reported for the case of MoS$_2$ under 6 ns, 532 nm, 10 Hz excitation conditions.[49] However, in the present experiments, due to the much lower laser intensities employed, the observed NLS was found to be insignificant for the range of laser intensities used, as it was verified experimentally. In fact, the much higher intensities used in ref. [28] and the higher repetition rate of the laser in ref. [49], necessarily entailed an increase in the thermal energy accumulated in the absorbing medium, thus facilitating the formation of scattering centers.

In Figure 7, some characteristic "divided" Z-scans of 2H-MoS$_2$, 1T-MoS$_2$, 2H-WS$_2$, and 1T-WS$_2$ are presented, obtained using visible (Figure 7a) and infrared (Figure 7b) laser excitation, respectively. The concentration of all the dispersions was 0.1 mg/mL. Since the NLO refraction of the solvents was found to be negligible for the incident laser intensities used (i.e., from 8.4 to 528 MW/cm$^2$), the shown "divided" Z-scans reveal straightforwardly the sign of the NLO refractive response of the TMDCs nanosheets. As shown, the 2H-MoS$_2$ and 2H-WS$_2$ were found to exhibit a valley-peak configuration, suggesting self-focusing action (i.e., $\gamma' > 0$), while the 1T-MoS$_2$ and 1T-WS$_2$ (i.e., the metallic phase of MoS$_2$ and WS$_2$), were found exhibiting negative NLO refractive response, as denoted by the peak-valley shape, suggesting self-defocusing action (i.e., $\gamma' < 0$). The same behavior was observed for both excitation wavelengths.

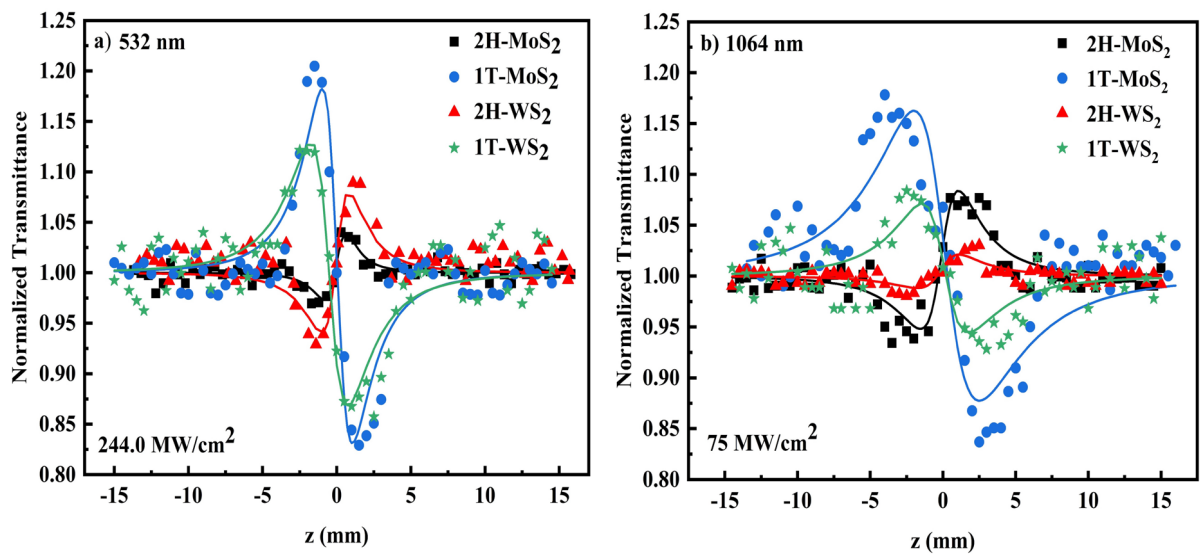

**Figure 7.** "Divided" Z-scans of 2H-MoS$_2$, 1T-MoS$_2$, 2H-WS$_2$, 2H/1T-WS$_2$ and 1T-WS$_2$ dispersions, under (a) 532 and (b) 1064 nm, all having a concentration of 0.1 mg/mL.

The physical origins of the nonlinear refraction of the studied TMDCs can be attributed to the instantaneous bound-electronic response (Kerr nonlinearity) and also to other non-instantaneous contributions, such as molecular reorientation, free carrier refraction and thermal effects.[50] As discussed previously, 1PA or 2PA processes can result in excitation of electrons from the valence band to the conduction band, generating free carriers. For longer laser excitation pulses, e.g., for ns pulses, due to the more efficient electronic transitions occurring, free carrier refraction is the most sizable contribution to the observed refractive nonlinearity, overwhelming the (faster) response of the bound electronic and molecular reorientation nonlinearities.[48]

In addition, the manifestation of thermal effects for the case of 2H-MoS$_2$ and 2H-WS$_2$ can be excluded, since, in principle, thermal effects result in self-defocusing behavior.[51] Therefore, only the NLO refraction of 1T-MoS$_2$ and 1T-WS$_2$ could be related to a thermal origin nonlinearity. However, according to the literature,[52] in the case of a thermal nonlinearity, the value of Δz (defined as the distance between the peak and the valley of the "divided" Z-scan), is usually related to $z_0$ (i.e., $z_0$ is the Rayleigh length) through the relation: Δz ≈ 1.1$z_o$. In the present measurements, the Rayleigh length was determined to be 1.91 and 2.65 mm for the 532 and 1064 nm laser beams, respectively, while the corresponding Δz values, as determined form the "divided" Z-scans, were 2.8 and 6 mm respectively. So, it can be safely assumed that the dominant mechanism contributing to the observed NLO refraction of 2H-MoS$_2$, 1T-MoS$_2$, 2H-WS$_2$ and 1T-WS$_2$ can be most probably attributed to free carrier refraction from direct one-photon or two-photon optical transitions.

From the analysis of the data of the Z-scan experiments following the procedure described in detail elsewhere,[30] the NLO parameters of the trigonal prismatic and octahedral phases of MoS$_2$ and WS$_2$ nanosheets, namely the nonlinear absorption coefficient β, the nonlinear refractive index parameter γ′ and the third-order nonlinear susceptibility χ$^{(3)}$ have been determined and are summarized in Table 1. To facilitate comparisons, all reported values are referring to a concentration of 1 mg/mL.

**Table 1.** Determined NLO parameters of 2H-MoS$_2$, 1T-MoS$_2$, 2H-WS$_2$ and 1T-WS$_2$ (all values referring to a concentration of 1 mg/mL).

| λ (nm) | Sample* | β (×10$^{-11}$ m/W) | γ' (×10$^{-18}$ m$^2$/W) | \|χ\|$^{(3)}$ (×10$^{-13}$ esu) |
|---|---|---|---|---|
| 532 | 1T-MoS$_2$ | -919.8 ± 96.0 | -759.4 ± 103.0 | 953.8 ± 125.0 |
| | 2H-MoS$_2$ | 166.6 ± 17 | 256 ± 34 | 348.1 ± 48.0 |
| | 1T-WS$_2$ | -287.7 ± 36.0 | -613 ± 45 | 701 ± 59 |
| | 2H-WS$_2$ | 131.1 ± 14.0 | 424.4 ± 60.0 | 554.2 ± 78.0 |
| 1064 | 1T-MoS$_2$ | -291.1 ± 26.0 | -245.0 ± 24.0 | 388.9 ± 24 |
| | 2H-MoS$_2$ | 115.1 ± 13 | 131.4 ± 12.0 | 219.4 ± 13.0 |
| | 1T-WS$_2$ | -97.1 ± 10.0 | -128 ± 18.0 | 172 ± 12 |
| | 2H-WS$_2$ | 31.2 ± 4.0 | 71.5 ± 9.0 | 99.4 ± 13 |

*1T-MoS$_2$ and 1T-WS$_2$ were dispersed in H$_2$O, the 2H-MoS$_2$ and 2H-WS$_2$ were dispersed in DMF.

As can be seen from the values of the NLO parameters of Table 1, the present findings confirm undoubtedly that the coordination of Mo and W atoms (i.e., trigonal prismatic or octahedral) affects drastically both the sign and the magnitude of the NLO response of MoS$_2$ and WS$_2$. More specifically, the 1T-metallic MoS$_2$ and WS$_2$ exhibited enhanced NLO response compared to their 2H-semiconducting counterparts, at both excitation wavelengths. This is expressed by the larger third-order susceptibility χ$^{(3)}$ of the 1T-metallic MoS$_2$ and WS$_2$ which is larger by a factor of about 3 and 1.5, respectively, under visible excitation regime, and by a factor of ~2, under infrared excitation compared to the 2H-semiconducting MoS$_2$ and WS$_2$ structures. The above findings clearly reveal the importance of the crystalline phase of MoS$_2$ and WS$_2$ (i.e., metallic or semiconducting) as an efficient tool for the enhancement and tailoring of their NLO response, in view of optoelectronic and photonic applications.

Concerning the NLO absorptive response of the investigated TMDCs, as can be seen from the nonlinear absorption coefficient β values listed in Table 1, two general observations can be drawn: the metallic phase TDMCs exhibited significantly larger NLO absorption than the semiconducting phase ones, and that the Mo containing TDMCs exhibited larger β values than their W-based counterparts. More specifically, the 1T-MoS$_2$ exhibited 6 and 2.5 times larger β values than the 2H-MoS$_2$ under visible and infrared laser excitation, respectively. In the same direction, 1T-WS$_2$ exhibited 2 and 3 times larger β than 2H-WS$_2$ at each excitation wavelength, respectively. It is worth to emphasized at this point that the larger β of 2H-MoS$_2$ compared to that of 2H-WS$_2$, under both excitation wavelengths, suggests a larger 2PA absorption cross section for the former. In addition, the metallic character TDMCs, 1T-MoS$_2$ and 1T-WS$_2$, exhibiting saturable absorption behavior (i.e., β<0) even for the highest laser intensities used, suggest a weaker 2PA absorption cross section, revealing their potential use as efficient saturable absorbers.

Concerning the nonlinear refractive response of the studied TDMCs, the metallic phase samples exhibited larger nonlinear refractive index parameter γ′ values than their semiconducting phase counterparts, while all samples exhibited larger γ′ for visible excitation. Interestingly, the metallic phase TDMCs 1T-MoS$_2$ and 1T-WS$_2$ were found to exhibit negative sign nonlinear refraction (i.e., γ′<0), corresponding to self-defocusing, while the semiconducting phase 2H-MoS$_2$ and 2H-WS$_2$ exhibited clearly positive nonlinear refraction (i.e., γ′>0), corresponding to self-focusing. It should be emphasized that all TDMCs exhibited large nonlinear refractive nonlinearity (i.e., of the order of $10^{-16}$ m$^2$/W,) which makes them interesting candidates for all-optical switching applications.

At this point, the role of defects should be discussed. The defects are inevitably introduced during the fabrication process of 2H-MoS$_2$ and 2H-WS$_2$ nanosheets. So, according to the literature, the defects most commonly observed in single- or few-layered semiconducting MoS$_2$ and WS$_2$, include S vacancies, Mo (or W) vacancies, S$_2$ double vacancies, a vacancy complex of Mo (or W) and three nearby sulfurs, a vacancy complex of Mo (or W) and three nearby disulfur pairs, and antisite defects where a Mo (or W) atom replaces a S atom, or vice versa.[53] The presence of these defects is expected to modify the electronic properties of MoS$_2$ and WS$_2$. For instance, the vacancies on the surface of the sheet behave as strong electron acceptors, inducing p-type characteristics in MoS$_2$ and WS$_2$. On the other hand, antisite defects act as electron donors, provide n-type doping behavior. Therefore, the coexistence of vacancies and antisites in the lattice of 2H-MoS$_2$ and 2H-WS$_2$, induce a push-pull-like system contributing to charge transfer effects in the sheet. In addition, after Li intercalation, the 2H-MoS$_2$ and 2H-WS$_2$ are subjected a phase transformation to 1T-MoS$_2$ and 1T-WS$_2$ lattice structure, respectively, giving rise to metallic character. In addition, it has been shown that during the lithiation process the number of structural defects is progressively increasing,[54] resulting in more efficient charge transfer in the sheet. In the present case, the increase of defects after Li intercalation can be corroborated by the broadening of the A$_{1g}$ and E$_{2g}^1$ modes and the appearance of several other defect-activated peaks, as can be seen in the Raman spectra of Figure 2. As the Li atoms are adopted in the non-defective 2H-MoS$_2$ and 2H-WS$_2$ sheets, there are two possible adsorption positions of the Li atoms, namely a top site above a Mo or W atom, and the hollow site in the hexagonal center. However, in presence of vacancies, the adsorption position is strongly dependent on the type of vacancy, as it has been predicted according to first principle calculations.[55] In all cases the incorporation of Li atoms results in a n-type doping state,[56] denoting the formation of an efficient charge transfer system. A similar behavior, i.e., charge transfer from Li-atoms to the surface of the sheet, has been also reported for other 2DMLs, such as graphene and silicene.[57,58] Therefore, the 1T phase of $MoS_2$ and $WS_2$ is expected to exhibit more efficient charge transfer compared to 2H phase, due to the presence of larger amount of point defects and the adsorption of Li atoms as well. However, it is known that systems with more efficient charge transfer exhibit, usually, stronger NLO response. This is in accordance with the present findings where the 1T-$MoS_2$ and 1T-$WS_2$ samples were found to exhibit the largest $\chi^{(3)}$.

Finally, the optical limiting (OL) action of the present TMDCs was studied in the spectral region from 450 to 1750 nm, by irradiating their dispersions with different wavelength laser radiations. For the experiments, dispersions of each TMDC were prepared having the same linear transmittance, i.e., ~70%, at each different irradiation wavelength, to facilitate comparisons. During these experiments, the fluence of the incident laser radiation, $F_{in}$, was varied and the fluence of the transmitted laser radiation, $F_{out}$, just after the sample, was measured. Both $F_{in}$ and $F_{out}$ were monitored simultaneously by two identical joulemeters. For the quantification of the OL performance of each sample, the value of the optical limiting onset, $OL_{on}$, is used, which is defined as the value of the incident fluence, $F_{in}$, at which sample's transmittance starts to deviate from the Beer-Lambert law. For the irradiation of the dispersions both the outputs of a Nd:YAG laser (i.e., the fundamental at 1064 nm and the SHG at 532 nm), and the tunable output of an optical parametric oscillator (OPO) pumped by the same Nd:YAG laser were used.

From these measurements, the semiconducting phase 2H-$MoS_2$ and 2H-$WS_2$ samples were found exhibiting strong and broadband OL action, while the metallic phase 1T-$MoS_2$ and 1T-$WS_2$ samples did not present any OL for all laser irradiation conditions used (from Vis to NIR), suggesting a weak or negligible two-photon absorption cross section. The OL performance of the 2H-$MoS_2$ and 2H-$WS_2$ dispersions under 532 and 1064 nm is presented in Figure 8. The dotted (straight) lines correspond to the sample's linear transmittance, which was ~ 70% for each excitation wavelength, while the solid lines (connecting the experimental data points) are for guiding the eye. From these curves, the $OL_{on}$ values of 2H-$MoS_2$ and 2H-$WS_2$ were determined to be about 0.92 and 1.07 $J/cm^2$, respectively, under 532 nm irradiation, and 0.9 and 1.15 $J/cm^2$ under 1064 nm irradiation.

In another study, performed under very similar experimental conditions, i.e., low repetition rate (2 Hz) 6 ns, 532 and 1064 nm laser excitation, and studying some $MoS_2$ and $WS_2$ nanosheets dispersions having similar linear transmittances (i.e., 55-80%), among other TDMCs, the reported $OL_{on}$ values were larger in general, but of the same order of magnitude, to the ones determined here.[28] The results of the two studies are in good agreement.

Comparing the $OL_{on}$ values of the present TDMCs for visible laser radiations (i.e., 532 nm) with that of an identical transmittance (i.e., ~70%) fullerene-$C_{60}$ solution, the latter presents lower $OL_{on}$ values, i.e., of ~0.16 J/cm$^2$, than the present TDMCs.[22] It is reminded that $C_{60}$ solutions are often used for comparison purposes in optical limiting studies for visible laser radiations. Unfortunately, no comparison can be done for NIR wavelengths, since as it known, fullerene-$C_{60}$ solutions exhibit negligible optical limiting in the infrared.[59]

In Figures S(2-4), the optical limiting performance of the 2H-MoS$_2$ and 2H-WS$_2$ under different visible and NIR irradiation wavelengths is shown. To the best of our knowledge, this is the first comprehensive study of the optical limiting of MoS$_2$ and WS$_2$ covering such a wide spectral range, while simultaneously aiming to elucidate the effect of crystalline phase on the broadband OL performance of these TMDCs. As shown, the values of the OL$_{on}$ were found to vary with the wavelength of the laser irradiation. More importantly, it was found that the onset of optical limiting action attained very low values in the NIR region, being at least one order of magnitude lower than the corresponding OL$_{on}$ values determined under visible irradiation. All determined OL$_{on}$ values of 2H-MoS$_2$ and 2H-WS$_2$ are summarized in Table 2. The present findings are of essential importance as they demonstrate the broadband OL performance of MoS$_2$ and WS$_2$, but also reveal their great OL efficiency in the NIR region, rendering them attractive candidates for optical limiting applications, such as protection of delicate optical components and human retinal, shaping and compression of an optical pulse, etc.[60,61]

**Table 2.** OL$_{on}$ values of 2H-MoS$_2$ and 2H-WS$_2$ under different irradiation wavelength.

| Wavelength (nm) | OL$_{on}$ (Jcm$^{-2}$) 2H-MoS$_2$ | OL$_{on}$ (Jcm$^{-2}$) 2H-WS$_2$ |
|---|---|---|
| 450 | 1.08 ± 0.06 | 1.24 ± 0.08 |
| 532 | 0.92 ± 0.06 | 1.07 ± 0.06 |
| 650 | 0.95 ± 0.08 | 1.03 ± 0.05 |
| 750 | 0.80 ± 0.07 | 1.0 ± 0.1 |
| 850 | 0.8 ± 0.05 | 1.05 ± 0.07 |
| 950 | 0.76 ± 0.06 | 1.16 ± 0.09 |
| 1064 | 0.90 ± 0.06 | 1.15 ± 0.08 |
| 1150 | 0.58 ± 0.06 | 0.75 ± 0.08 |
| 1250 | 0.32 ± 0.05 | 0.43 ± 0.04 |
| 1350 | 0.28 ± 0.07 | 0.32 ± 0.08 |
| 1450 | 0.20 ± 0.02 | 0.23 ± 0.03 |
| 1550 | 0.12 ± 0.02 | 0.14 ± 0.02 |
| 1650 | 0.11 ± 0.03 | 0.17 ± 0.04 |
| 1750 | 0.09 ± 0.01 | 0.11 ± 0.01 |

An overview of the variation of the $OL_{on}$ values of $2H$-$MoS_2$ and $2H$-$WS_2$ dispersions with the laser irradiation wavelength, from 450 to 1750 nm, is presented in Figure 8. As shown, the studied TMDCs exhibited a moderate OL performance for wavelengths up to 1064 nm, increasing importantly toward longer NIR wavelengths, where the $OL_{on}$ were determined to reach low values of about 0.09 and 0.11 for $2H$-$MoS_2$ and $2H$-$WS_2$, respectively. In all cases, 2H-semiconducting $MoS_2$ was found exhibiting more efficient optical limiting compared to the 2H-semiconducting phase of $WS_2$.

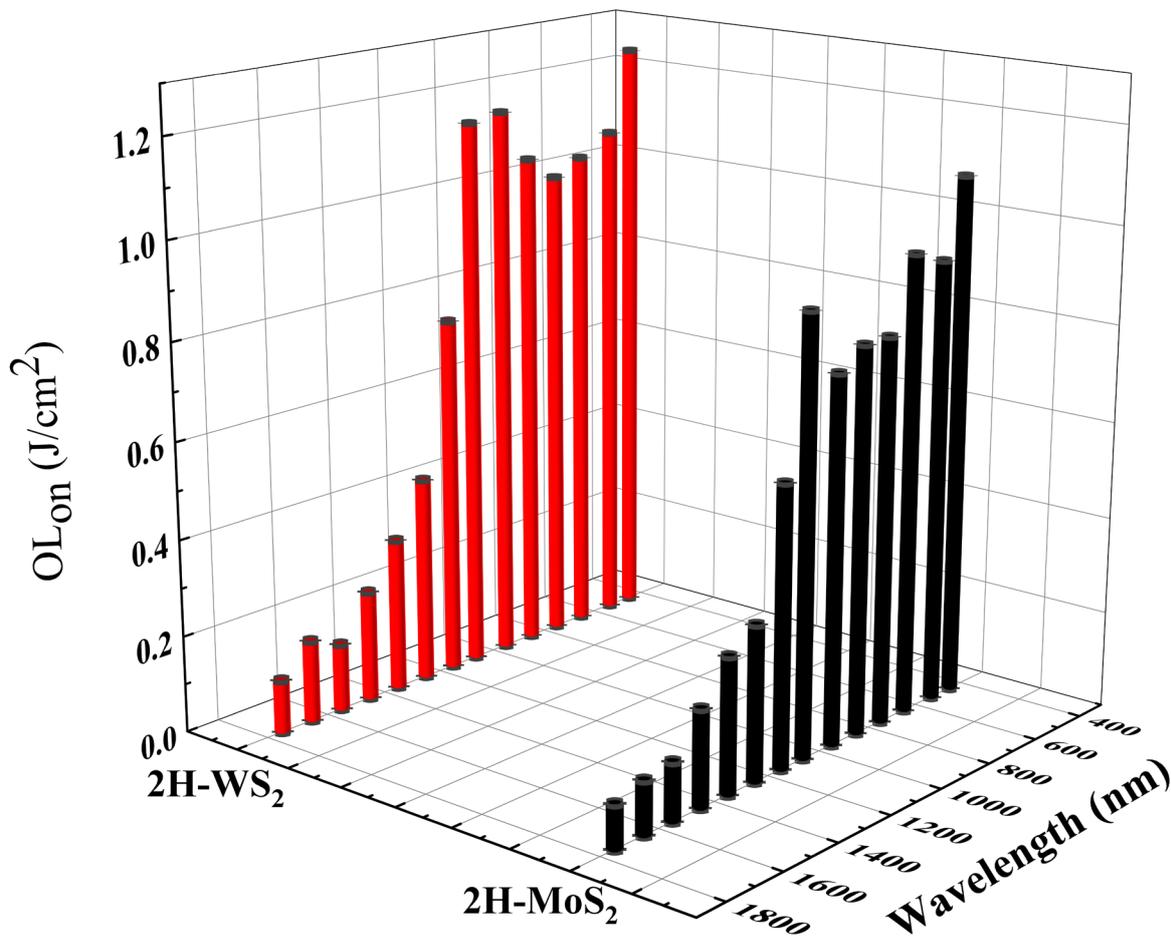

**Figure 8.** Variation of the optical limiting onset values, $OL_{on}$, of $2H$-$MoS_2$ and $2H$-$WS_2$, versus the laser irradiation wavelength.

There are several physical processes that can contribute to the observed OL action of the studied TMDCs, under nanosecond laser excitation. Among them, the most important are TPA/MPA (two-, multi-photon absorption), excited state absorption (ESA), free carrier absorption, nonlinear scattering (NLS), and thermal cumulative effects. The presence of NLS was checked and it was found to be insignificant under the present experimental conditions. Concerning thermal cumulative effects, any contribution arising from them can be excluded, as OL and Z-scan experiments performed under various repetition rates of the excitation laser did not show any variation of the observed OL action. In addition, the laser repetition rate was kept as low as 1 Hz. Hence, the OL performance of the 2H-semiconducting $MoS_2$ and $WS_2$ throughout the visible and NIR spectral regions should be ascribed to TPA/MPA and/or ESA, and free carrier absorption processes. The latter process has been also suggested recently, in a work combining "open aperture" Z-scan and photoacoustic Z-scan, providing a useful distinction between the nonlinear scattering and absorption effects.[62]

Next, for a more complete assessment of the OL performance of the present TMDCs, their $OL_{on}$ values will be compared to that of some other 2D nanostructures studied under similar experimental conditions and having attracted recently the research interest. In that view, the OL performance of the present 2DLMs will be compared to that of some other types of TMDCs, including $MoSe_2$, $WSe_2$, $TiS_2$, and $SnX_2$ (where X=S, Se) nanosheets,[28,49] some graphene nanosheets,[63] 2D black phosphorus nanosheets,[64] antimonene nanosheets,[65] and some vanadium carbide ($V_2CTx$) nanosheets.[66] Specifically, the $OL_{on}$ values of $MoSe_2$ and $WSe_2$ nanosheets were reported to be ~1.47 and 0.99 J cm$^{-2}$, respectively, at 532 nm and ~1.37 and 2.3 J cm$^{-2}$ at 1064 nm, corresponding to a lower efficiency of optical limiting compared to the presently studied TMDCs. On the other hand, the $OL_{on}$ values of $TiS_2$ and $SnX_2$ were found to be almost one order of magnitude lower than those of $MoS_2$ and $WS_2$ at 532 nm. Concerning the OL of graphene nanosheets dispersions, the values of $OL_{on}$ were reported to be greater than ~0.5 and 1 J cm$^{-2}$ at 532 and 1064 nm, respectively, corresponding to relatively better OL action for visible radiations and similar OL efficiency for infrared radiations. The present TMDCs were also found to exhibit similar or relatively better OL action than antimonene nanosheets, as the latter present $OL_{on}$ values greater than 0.8 J cm$^{-2}$. Concerning the OL of black phosphorus nanosheets dispersions, a $OL_{on}$ of ~10 J/cm$^2$ was reported, which is much larger than that of the present TMDCs. Similarly, in the case of vanadium carbide ($V_2CTx$) nanosheets, it was reported that their OL started to appear for laser fluences greater than 4 J/cm$^2$. It should be emphasized that, in all cases, the OL efficiency of both semiconducting $MoS_2$ and $WS_2$ to longer NIR wavelengths was found to be comparable or significantly better than that reported for the aforementioned 2DLMs under 532 and 1064 nm laser excitation. However, although the present TMDCs exhibit very low

OL$_{on}$ values at longer NIR wavelengths, a full comparison is difficult, as no available data exist for the other materials at this wavelength range.

**CONCLUSIONS**

In summary, in the present work some MoS$_2$ and WS$_2$ nanosheets were prepared and characterized by HRS-TEM microscopy, Raman spectroscopy and UV-Vis-NIR absorption spectroscopy. By choosing suitable preparation procedures, semiconducting (2H-MoS$_2$ and 2H-WS$_2$) and metallic (1T-MoS$_2$ and 1T-WS$_2$) phase MoS$_2$ and WS$_2$ nanosheets were obtained. Then, their third order NLO response was investigated under 4 ns, 532 and 1064 nm laser excitation. The experimental findings have shown that their NLO response depends strongly on their crystalline phase. More specifically, the NLO response of 1T-metallic MoS$_2$ and WS$_2$ nanosheets was found larger than that of their 2H-semiconducting counterparts, most probably due to the increased number of structural defects and the adsorption of Li atoms after Li intercalation, resulting in the formation of an efficient charge transfer system. Moreover, the 2H-MoS$_2$ and 2H-WS$_2$ were found to exhibit SA behavior at low laser intensities switching to RSA at higher laser intensities due to 2PA process, while their 1T-phase revealed a strong SA behavior even for the highest laser intensities employed. Notably, the semiconducting and metallic phase MoS$_2$ and WS$_2$ nanosheets were found exhibiting opposite sign refractive nonlinearities, i.e., self-focusing for 2H-semiconducting MoS$_2$ and WS$_2$, and self-defocusing for 1T-metallic MoS$_2$ and WS$_2$, due to strong free carrier refraction from direct one-photon or two-photon optical transitions. In addition, the OL action of the present nanosheets was studied by using different irradiation wavelengths ranging from 450 to 1750 nm. In particular, the 2H-semiconducting MoS$_2$ and WS$_2$ were found to exhibit exceptionally broadband OL action, attaining very low values of optical limiting onset toward NIR wavelengths, while the 1T-metallic MoS2 and WS2 did not show any OL action. The present findings demonstrated unambiguously that the NLO properties and the optical limiting action of these TDMCs nanosheets can be effectively controlled through their crystalline phase, thus offering an efficient strategy for preparing TMDCs with custom made NLO properties for various optoelectronic and photonic applications, as e.g., optical limiting, saturable absorption and optical switching.

**Notes**

The authors declare no competing financial interests.

**ACKNOWLEDGMENT**

M.S. acknowledges support from the Hellenic Foundation for Research and Innovation (HFRI) under the HFRI PhD Fellowship grant (number: 83656). Financial support to N.T. by the Hellenic Foundation for Research and Innovation HFRI under the "2nd Call for HFRI Research Projects to support Faculty Members and Researchers" (Project Number: 2482) is acknowledged. TEM studies were performed in Laboratorio de Microscopias Avanzadas (LMA), Universidad de Zaragoza (Spain). R.A. acknowledges support from Spanish MICINN (PID2019-104739GB–100/AEI/10.13039/501100011033), Government of Aragon (projects DGA E13-20R) and from EU H2020 "ESTEEM3" (Grant number 823717) and Graphene Flagship (Grant number 881603).

**ASSOCIATED CONTENT**

**Supporting Information**

Raman measurements; NLO parameters of different concentrations of TMDCs; Optical limiting measurements at different irradiation wavelengths.

# REFERENCES

[1] Novoselov, K. S.; Geim, A. K.; Morozov, S. V.; Jiang, D.; Zhang, Y.; Dubonos, S. V.; Grigorieva, I. V.; Firsov, A. A. Electric Field Effect in Atomically Thin Carbon Films. *Science* **2004**, *306*, 666-669.

[2] Wang, L.; Meric, I.; Huang, P. Y.; Gao, Q.; Gao, Y.; Tran, H.; Taniguchi, T.; Watanabe, K.; Campos, L. M.; Muller, D. A.; Guo, J.; Kim, P.; Hone, J.; Shepard, K. L.; Dean, C. R. One-Dimensional Electrical Contact to a Two-Dimensional. *Science* **2013**, *342*, 614-617.

[3] Ni, G. X.; Wang, L.; Goldflam, M. D.; Wagner, M.; Fei, Z.; McLeod, A. S.; Liu, M. K.; Keilmann, F.; Özyilmaz, B.; Castro Neto, A. H.; Hone, J.; Fogler, M. M.; Basov, D. N. Ultrafast Optical Switching of Infrared Plasmon Polaritons in High-Mobility Graphene. *Nat. Photonics* **2016**, *10*, 244-247.

[4] Wang, Q. H.; Kalantar-Zadeh, K.; Kis, A.; Coleman, J. N.; Strano, M. S. Electronics and Optoelectronics of Two-Dimensional Transition Metal Dichalcogenides. *Nat. Nanotechnol.* **2012**, *7*, 699-712.

[5] Balis, N.; Stratakis, E.; Kymakis, E. Graphene and Transition Metal Dichalcogenide Nanosheets as Charge Transport Layers for Solution Processed Solar Cells. *Mater. Today* **2016**, *19*, 580-594.

[6] Park, Y.; Ryu, B.; Oh, B.-R.; Song, Y.; Liang, X.; Kurabayashi, K. Biotunable Nanoplasmonic Filter on Few-Layer $MoS_2$ for Rapid and Highly Sensitive Cytokine Optoelectronic Immunosensing. *ACS Nano* **2017**, *11*, 5697-5705.

[7] Li, B.; Zu, S.; Zhou, J.; Jiang, Q.; Du, B.; Shan, H.; Luo, Y.; Liu, Z.; Zhu, X.; Fang, Z. Single-Nanoparticle Plasmonic Electro-Optic Modulator Based on $MoS_2$ Monolayers. *ACS Nano* **2017**, *11*, 9720-9727.

[8] Zhang, H.; Lu, S. B.; Zheng, J.; Du, J.; Wen, S. C.; Tang, D. Y.; Loh, K. P. Molybdenum Disulfide ($MoS_2$) as a Broadband Saturable Absorber for Ultra-Fast Photonics. *Opt. Express* **2014**, *22*, 7249-7260.

[9] Tsai, D.-S.; Liu, K.-K.; Lien, D.-H.; Tsai, M.-L.; Kang, C.-F.; Lin, C.-A.; Li, L.-J.; He, J.-H. Few-Layer $MoS_2$ with High Broadband Photogain and Fast Optical Switching for Use in Harsh Environments. *ACS Nano* **2013**, *7*, 3905-3911.

[10] Liu, T.; Wang, C.; Gu, X.; Gong, H.; Cheng, L.; Shi, X.; Feng, L.; Sun, B.; Liu, Z. Drug Delivery with PEGylated $MoS_2$ Nano-Sheets for Combined Photothermal and Chemotherapy of Cancer. *Adv. Mater.* **2014**, *26*, 3433-3440.

[11] Han, J.; Xia, H.; Wu, Y.; Kong, S. N.; Deivasigamani, A.; Xu, R.; Hui, K. M.; Kang, Y. Single-Layer $MoS_2$ Nanosheet Grafted Upconversion Nanoparticles for Near-Infrared Fluorescence Imaging-Guided Deep Tissue Cancer Phototherapy. *Nanoscale* **2016**, *8*, 7861-7865.

[12] Zhu, C.; Zeng, Z.; Li, H.; Li, F.; Fan, C.; Zhang, H. Single-Layer $MoS_2$-Based Nanoprobes for Homogeneous Detection of Biomolecules. *J. Am. Chem. Soc.* **2013**, *135*, 5998–6001.

[13] Farimani, A. B.; Min, K.; Aluru, N. R. DNA Base Detection Using a Single-Layer $MoS_2$. *ACS Nano* **2014**, *8*, 7914-7922.

[14] Chhowalla, M.; Shin, H. S.; Eda, G.; Li, L.-J.; Loh, K. P.; Zhang, H. The Chemistry of Two-Dimensional Layered Transition Metal Dichalcogenide Nanosheets. *Nat. Chem.* **2013**, *5*, 263–275.

[15] Py, M. A.; Haering, R. R. Structural Destabilization Induced by Lithium Intercalation in $MoS_2$ and Related-Compounds. Can. J. Phys. **1983**, *61*, 76-84.

[16] Wang, H.; Lu, Z.; Kong, D.; Sun, J.; Hymel, T. M.; Cui, Y. Electrochemical Tuning of $MoS_2$ Nanoparticles on Three-Dimensional Substrate for Efficient Hydrogen Evolution. *ACS Nano* **2014**, *8*, 4940-4947.

[17] Splendiani, A.; Sun, L.; Zhang, Y.; Li, T.; Kim, J.; Chim, C.-Y.; Galli, G.; Wang, F. Emerging Photoluminescence in Monolayer $MoS_2$. *Nano Lett.* **2010**, *10*, 1271-1275.

[18] Mak, K. F.; Lee, C.; Hone, J.; Shan, J.; Heinz, T. F. Atomically Thin $MoS_2$: A New Direct-Gap Semiconductor. *Phys. Rev. Lett.* **2010**, *105*, 136805.

[19] Pang, L.; Liu, W.; Tian, W.; Han, H.; Wei, Z. Nanosecond Hybrid *Q*-Switched Er-Doped Fiber Laser with $WS_2$ Saturable Absorber. *IEEE Photon. J.* **2016**, *8*, 1501907.

[20] Asobe, M.; Kanamori, T.; Kubodera, K. Ultrafast All-Optical Switching Using Highly Nonlinear Chalcogenide Glass Fiber. *IEEE Photon. Technol. Lett.* **1992**, *4*, 362-365.

[21] Langrock, C.; Kumar, S.; McGeehan, J. E.; Willner, A. E.; Fejer, M. M. All-Optical Signal Processing Using χ$^{(2)}$ Nonlinearities in Guided-Wave Devices. *J. Light. Technol.* **2006**, *24*, 2579-2592.

[22] Stavrou, M.; Papadakis, I.; Bawari, S.; Narayanan, T. N.; Couris, S. Giant Broadband (450–2300 nm) Optical Limiting and Enhancement of the Nonlinear Optical Response of Some Graphenes by Defect Engineering. *J. Phys. Chem. C* **2021**, *125*, 16075-16085.

[23] Papadakis, I.; Stavrou, M.; Bawari, S.; Narayanan, T. N.; Couris, S. Outstanding Broadband (532 nm to 2.2 μm) and Very Efficient Optical Limiting Performance of Some Defect-Engineered Graphenes. *J. Phys. Chem. Lett.* **2020**, *11*, 9515-9520.

[24] Stathis, A.; Stavrou, M.; Papadakis, I.; Obratzov, I.; Couris, S. Enhancing and Tuning the Nonlinear Optical Response and Wavelength-Agile Strong Optical Limiting Action of N-octylamine Modified Fluorographenes. *Nanomaterials* **2020**, *10*, 2319.

[25] Bao, Q.; Zhang, H.; Wang, Y.; Ni, Z.; Yan, Y.; Shen, Z. X.; Loh, K. P.; Tang, D. Y. Atomic-Layer Graphene as a Saturable Absorber for Ultrafast Pulsed Lasers. *Adv. Funct. Mater.* **2009**, *19*, 3077-3083.

[26] Wang, K.; Wang, J.; Fan, J.; Lotya, M.; O'Neill, A.; Fox, D.; Feng, Y.; Zhang, X.; Jiang, B.; Zhao, Q., et al. Ultrafast Saturable Absorption of Two-Dimensional $MoS_2$ Nanosheets. *ACS Nano* **2013**, *7*, 9260-9267.

[27] Wang, K.; Feng, Y.; Chang, C.; Zhan, J.; Wang, C.; Zhao, Q.; Coleman, J. N.; Zhang, L.; Blau, W. J.; Wang, J. Broadband Ultrafast Nonlinear Absorption and Nonlinear Refraction of Layered Molybdenum Dichalcogenide Semiconductors. *Nanoscale* **2014**, *6*, 10530-10535.

[28] Dong, N.; Li, Y.; Feng, Y.; Zhang, S.; Zhang, X.; Chang, C.; Fan, J.; Zhang, L.; Wang, J. Optical Limiting and Theoretical Modelling of Layered Transition Metal Dichalcogenide Nanosheets. *Sci. Rep.* **2015**, *5*, 14646.

[29] Zhou, K.-G.; Zhao, M.; Chang, M.-J.; Wang, Q.; Wu, X.-Z.; Song, Y.; Zhang, H.-L. Size-Dependent Nonlinear Optical Properties of Atomically Thin Transition Metal Dichalcogenide Nanosheets. *Small* **2014**, *11*, 694-701.

[30] Sheik-bahae, M.; Said, A. A.; Van Stryland, E. W. High-Sensitivity, Single-Beam $n_2$ Measurements. *Opt. Lett.* **1989**, *14*, 955-957.

[31] Papagiannouli, I.; Iliopoulos, K.; Gindre, D.; Sahraoui, B.; Krupka, O.; Smokal, V.; Kolendo, A.; Couris, S. Third-order Nonlinear Optical Response of Push-Pull Azobenzene Polymers. *Chem. Phys. Lett.* **2012**, *554*, 107-112.

[32] Pagona, G.; Bittencourt, C.; Arenal, R.; Tagmatarchis, N. Exfoliated Semiconducting Pure 2H-$MoS_2$ and 2H-$WS_2$ Assisted by Chlorosulfonic Acid. *Chem. Commun.* **2015**, *51*, 12950.

[33] Chrissafis, K.; Zamani, M.; Kambas, K.; Stoemenos, J.; Economou, N. A.; Samaras, I.; Julien, C. Structural Studies of $MoS_2$ Intercalated by Lithium. *Mater. Sci. Eng., B,* **1989**, *3*, 145-151.

[34] Zhang, X.; Qiao, X. F.; Shi, W.; Wu, J. B.; Jiang D. S.; Tan, P. H. Phonon and Raman Scattering of Two-Dimensional Transition Metal Dichalcogenides from Monolayer, Multilayer to Bulk Material. *Chem. Soc. Rev.* **2015**, *44*, 2757–2785.

[35] Voiry, D.; Yamaguchi, H.; Li, J.; Silva, R.; Alves, D. C. B.; Fujita, T.; Chen, M.; Asefa, T.; Shenoy, V. B.; Eda, G., et al. Enhanced Catalytic Activity in Strained Chemically Exfoliated $WS_2$ Nanosheets for Hydrogen Evolution. *Nat. Mater*. **2013**, *12*, 850-855.

[36] Pierucci, D.; Zribi, J.; Livache, C.; Gréboval, C.; Silly, M. G.; Chaste, J.; Patriarche, G.; Montarnal, D.; Lhuillier, E.; Ouerghi, A., et al. Evidence for A Narrow Band Gap Phase In 1T′ $WS_2$ Nanosheet. *Appl. Phys. Lett.* **2019**, *115*, 032102.

[37] Sideri, I. K., Arenal, R.; Tagmatarchis, N. Covalently Functionalized $MoS_2$ with Dithiolenes. *ACS Mater. Lett.* **2020**, *2*, 832–837.

[38] Wu, R. J.; Odlyzko, M. L.; Mkhoyan, K. A. Determining the Thickness of Atomically Thin MoS2 and WS2 in the TEM. *Ultramicroscopy* **2014**, *147*, 8–20.

[39] Serra, M.; Arenal, R.; Tenne, R. An overview of the recent advances in inorganic nanotubes. *Nanoscale* **2019**, *11*, 8073-8090.

[40] Coehoorn, R.; Haas, C.; de Groot, R. A.; Electronic Structure of $MoSe_2$, $MoS_2$, and $WSe_2$. II. The Nature of the Optical Band Gaps. *Phys. Rev. B* **1987**, *35*, 6203-6206.

[41] Gan, Z. X.; Liu, L. Z.; Wu, H. Y.; Hao, Y. L.; Shan, Y.; Wu, X. L.; Chu, P. K. Quantum Confinement Effects Across Two-Dimensional Planes in $MoS_2$ Quantum Dots. *Appl. Phys. Lett.* **2015**, *106*, 233113.

[39] Zhao, W.; Ghorannevis, Z.; Chu, L.; Toh, M.; Kloc, C.; Tan, P. H.; Eda, G. Evolution of Electronic Structure in Atomically Thin Sheets of WS$_2$ and WSe$_2$. *ACS Nano* **2013**, *7*, 791-797.

[43] Gao, Y.; Zhang, X.; Li, Y.; Liu, H.; Wang, Y.; Chang, Q.; Jiao, W.; Song, Y. Saturable Absorption and Reverse Saturable Absorption in Platinum Nanoparticles. *Opt. Commun.* **2005**, *251*, 429-433.

[44] Lami, J.-F.; Gilliot, P.; Hirlimann, C. Observation of Interband Two-Photon Absorption Saturation in CdS. *Phys. Rev. Lett.* **1996**, *77*, 1632.

[45] Shi, H.; Yan, R.; Bertolazzi, S.; Brivio, J.; Gao, B.; Kis, A.; Jena, D.; Xing, H. G.; Huang, L. Exciton Dynamics in Suspended Monolayer and Few-Layer MoS$_2$ 2D Crystals. *ACS Nano* **2013**, *7*, 1072-1080.

[46] Goswami, T.; Bhatt, H.; Babu, K. J.; Kaur, G.; Ghorai, N.; Ghosh, H. N. Ultrafast Insights into High Energy (C and D) Excitons in Few Layer WS$_2$. *J. Phys. Chem. Lett.* **2021**, *12*, 6526-6534.

[47] Stavrou, M.; Papaparaskeva, G.; Stathis, A.; Stylianou, A.; Turcu, R.; Krasia-Christoforou, T.; Couris, S. Synthesis, Characterization and Nonlinear Optical Response of Polyelectrolyte-Stabilized Copper Hydroxide and Copper Oxide Colloidal Nanohybrids. *Opt. Mater.* **2021**, *119*, 111329.

[48] Said, A. A.; Sheik-Bahae, M.; Hagan, D. J.; Wei, T. H.; Wang, J.; Young, J.; Van Stryland, E. W. Determination of Bound-Electronic and Free-Carrier Nonlinearities in ZnSe, GaAs, CdTe, and ZnTe. *J. Opt. Soc. Am. B* **1992**, *9*, 405-414.

[49] Dong, N.; Li, Y.; Zhang, S.; Zhang, X.; Wang, J. Optically Induced Transparency and Extinction in Dispersed MoS$_2$, MoSe$_2$, and Graphene Nanosheets. *Adv. Opt. Mater.* **2017**, *5*, 1700543.]

[50] Sutherland, R. L. *Handbook of Nonlinear Optics*; Marcel Dekker: New York, 1998.

[51] Stavrou, M.; Dalamaras, I.; Karampitsos, N.; Couris, S. Determination of the Nonlinear Optical Properties of Single- and Few-Layered Graphene Dispersions under Femtosecond Laser Excitation: Electronic and Thermal Origin Contributions. *J. Phys. Chem. C* **2020**, *124*, 27241-27249.

[52] Falconieri, M.; Salvetti, G. Simultaneous Measurement of Pure-Optical and Thermo-Optical Nonlinearities Induced by High Repetition-Rate, Femtosecond Laser Pulses: Application to $CS_2$. *Appl. Phys. B* **1999**, *69*, 133-136.

[53] Komsa, H.-P.; Krasheninnikov, A. V. Native Defects in Bulk and $MoS_2$ Monolayer from First Principles. *Phys. Rev. B* **2015**, *91*, 125304.

[54] Xiong, F.; Wang, H.; Liu, X.; Sun, J.; Brongersma, M.; Pop, E.; Cui, Y. Li Intercalation in $MoS_2$: In Situ Observation of its Dynamics and Tuning Optical and Electrical Properties. *Nano Lett.* **2015**, *15*, 6777-6784.

[55] Sun, X.; Wang, Z.; Fu, Y. Q. Defect-Mediated Lithium Adsorption and Diffusion on Monolayer Molybdenum Disulfide. *Sci. Rep.* **2015**, *5*, 18712.

[56] Kaloni, T. P.; Kahaly, M. U.; Cheng, Y. C.; Schwingenschlögl, U. K-Intercalated Carbon Systems: Effects of Dimensionality and Substrate. *Europhys. Lett.* **2012**, *98*, 67003.

[57] Kaloni, T. P.; Schreckenbach, G.; Freund, M. S. Large Enhancement and Tunable Band Gap in Silicene by Small Organic Molecule Adsorption. *J. Phys. Chem. C* **2014**, *118*, 23361-23367.

[58] Denis, P. A. Chemical Reactivity of Lithium Doped Monolayer and Bilayer Graphene. *J. Phys. Chem. C* **2011**, *115*, 13392-13398.

[59] Couris, S.; Koudoumas, E.; Ruth, A. A.; Leach, S. Concentration and Wavelength Dependence of the Effective Third-Order Susceptibility and Optical Limiting of $C_{60}$ in Toluene Solution. *J. Phys. B: At. Mol. Opt. Phys.* **1995**, *28*, 4537.

[60] Spangler, C. W. Recent Development in the Design of Organic Materials for Optical Power Limiting. *J. Mater. Chem.* **1999**, *9*, 2013-2020

[61] Tutt, L. W.; Boggess, T. F. A Review of Optical Limiting Mechanisms and Devices Using Organics, Fullerenes, Semiconductors and Other Materials. *Prog. Quantum Electron.* **1993**, *17*, 299-338.

[62] Dhasmana, N.; Fadil, D.; Kaul, A. B.; Thomas, J. Investigation of Nonlinear Optical Properties of Exfoliated $MoS_2$ Using Photoacoustic Zscan. *MRS Adv.* **2016**, *1*, 3215-3221.

[63] Wang, J.; Hernandez, Y.; Lotya, M.; Coleman, J. N.; Blau, W. J. Broadband Nonlinear Optical Response of Graphene Dispersions. *Adv. Mater.* **2009**, *21*, 2430-2435.

[64] Huang, J.; Dong, N.; Zhang, S.; Sun, Z.; Zhang, W.; Wang, J. Nonlinear Absorption Induced Transparency and Optical Limiting of Black Phosphorus Nanosheets. *ACS Photonics* **2017**, *4*, 3063−3070.

[65] Shen, W.; Hu, J.; Ma, T.; Wang, J.; Wei, Y.; Zhang, Y.; Wu, J.; Chen, J. Antimonene Prepared by Laser Irradiation Applied for Nonlinear Optical Limiting. *Electron. Mater. Lett.* **2021**, *17*, 521-531.

[66] He, Q.; Hu, H.; Shao, Y.; Zhao, Z. Switchable Optical Nonlinear Properties of Monolayer $V_2CT_X$ MXene. *Optik* **2021**, *247*, 167629.